\documentclass[superscriptaddress,
reprint,longbibliography,
preprintnumbers,nofootinbib,
amsmath,amssymb,aps,prl]{revtex4-1}

\pdfoutput=1

\usepackage[utf8]{inputenc} 
\usepackage[T1]{fontenc} 
\usepackage{microtype}
\usepackage{mathrsfs}
\usepackage{amsmath}

\usepackage{graphicx}

\usepackage[dvipsnames]{xcolor}

\usepackage{dcolumn}
\usepackage{bm}

\usepackage{slashed}
\usepackage[colorlinks = true,
            linkcolor = blue,
            urlcolor  = blue,
            citecolor =red,
            anchorcolor = blue]{hyperref}
\usepackage{verbatim}
\usepackage{color,ulem}
\usepackage[english]{babel}

\usepackage[scr]{rsfso}

\def\be{\begin{equation}}
\def\ee{\end{equation}}
\def\ba{\begin{eqnarray}}
\def\ea{\end{eqnarray}}

\def\s{\sigma}

\newcommand{\prt}[1]{{\left( {#1} \right)}}

\def\IR{\relax{\rm I\kern-.18em R}}

\def\IR{\relax{\rm I\kern-.18em R}}
\def\IL{\relax{\rm I\kern-.18em L}}

\def\inv{^{\raise.15ex\hbox{${\scriptscriptstyle -}$}\kern-.05em 1}}

\def\bea{\begin{eqnarray}}
\def\eea{\end{eqnarray}}
\newcommand{\eq}[1]{(\ref{#1})}

\newcommand{\la}[1]{\label{#1}}

\def\s{\sigma}

\definecolor{markcolor2}{rgb}{1,0,0}

\definecolor{markcolor3}{rgb}{0,1,0}

\usepackage{blindtext}

\begin{document}

\preprint{IFT-UAM/CSIC-20-96}

\title{\Large Detecting Topological Quantum Phase Transitions via the c-Function}

\author{Matteo Baggioli}%
 \email{matteo.baggioli@uam.es}
\affiliation{Instituto de Fisica Teorica UAM/CSIC, c/Nicolas Cabrera 13-15,
Universidad Autonoma de Madrid, Cantoblanco, 28049 Madrid, Spain.}%

\author{Dimitrios Giataganas}%
 \email{dgiataganas@phys.uoa.gr}
\affiliation{Department of Physics,
University of Athens, 15771 Athens, Greece.}%
\affiliation{Physics Division, National Center for Theoretical Sciences,
National Tsing-Hua University, Hsinchu, 30013, Taiwan.}%


\begin{abstract}
We propose the c-function as a new and accurate probe to detect the location of topological quantum critical points. As a direct application, we consider a holographic model which exhibits a topological quantum phase transition between a topologically trivial insulating phase and a gapless Weyl semimetal. The quantum critical point displays a strong Lifshitz-like anisotropy in the spatial directions and the quantum phase transition does not follow the standard Landau paradigm. The c-function robustly shows a global \color{black}feature \color{black} at the quantum criticality  and distinguishes with great accuracy the two separate zero temperature phases. 
\color{black} Taking into account the relation of the c-function with the entanglement entropy, we \color{black} conjecture that our proposal is a general feature of quantum phase transitions and that is applicable beyond the holographic framework.  
\end{abstract}

\maketitle
\section{Introduction}
Phase transitions are ubiquitous in nature and they provide one of the most elegant examples of \textit{Universality} and a new window into the  physics of strongly correlated quantum many-body systems \cite{wen2004quantum}. Of exceptional interest are phase transitions happening at zero temperature -- the \textit{quantum phase transitions} \cite{sachdev2011quantum} -- which require a shift of paradigm within the condensed matter lore since they do not, \color{black}{in general}, \color{black} admit a simple Ginzburg-Landau description \cite{Landau:486430} and they are often not characterized by any spontaneous symmetry breaking pattern. A typical case is that of metal-insulator transitions \cite{metalinsulator_review}.\\ 

Topological quantum phase transitions (TQPT) are a particularly challenging subclass; the most famous example being quantum hall systems, displaying exotic features such as fractional statistics and topological degeneracy \cite{PhysRevLett.48.1559,PhysRevLett.50.1395}. The chase for an ''\textit{order parameter}'' or a local quantity able to pinpoint the location of the TQPT is a pressing and fundamental open question given the plethora of topological phases discovered in the recent years and their possible importance for technological developments such as quantum computing \cite{kitaev2009topological}.\\

In recent years, there have been several attempts to find an efficient observable able to locate the TQPT from the nature of the quasiparticles \cite{Manna2019QuasiparticlesAD}, the (not Ising-like) critical exponents  \cite{PhysRevLett.96.026802}, the \textit{dynamical topological order parameter} \cite{Xu2020} to other quantum information quantities such as \textit{fidelity} \cite{PhysRevA.78.010301} and \textit{topological entanglement entropy} \cite{PhysRevLett.96.110404,PhysRevLett.96.110405}. \\

In this work, we propose a different and particularly effective way to detect \color{black} the critical points  of \color{black} TQPTs by considering the \textit{c-function} of the system. We show that such quantity displays a neat and narrow signal at the location of the quantum critical point and it is therefore able to identify with precision the separation between the two topological phases.
\color{black} The c-function perfectly locates the position of the critical point even when the quantum transition is of topological nature, as it also does in continuous and discontinuous phase transitions.  \color{black} We also show that, independently of the microscopic details of the system such as the value of its quartic coupling, our c-function probe is still successful locating the transition. \color{black} 

The c-function is a natural candidate to detect phase transitions. The re-organization of the degrees of freedom (\textit{dofs}) along a transition is a key-feature to understand the two different phases involved. In relativistic theories, a clear measure for the number of effective degrees of freedom is indeed provided by the \textit{c-function}, whose monotonicity along the renormalization group (RG) flow is guaranteed by c-theorems \cite{Zamolodchikov:1986gt,Cardy:1988cwa,Komargodski:2011vj,Ryu:2006ef,Myers:2010tj,Casini:2004bw,Casini:2006es,Myers:2010xs}. These theorems formalize the idea that the number of \textit{dofs} diminishes monotonically flowing towards low energy and their validity is tightly connected with the existence of a null energy condition (NEC) \cite{Myers:2012ed}. At any fixed point, the \textit{c-function} coincides with the \textit{central charge} $c$ of the system, \color{black} which is again related to the theory's degrees of freedom. \color{black}\\ 

The proof of the c-theorems relies crucially on Lorentz invariance and the monotonicity of the \textit{c-function} is not guaranteed if such set of symmetries is broken \cite{Swingle:2013zla,Cremonini:2013ipa,Chu:2019uoh,Cremonini:2020rdx}. Additionally, when the rotational global symmetries are broken, as it happens in anisotropic Lifshitzs-like fixed points, the c-function needs to be redefined appropriately. Such a \textit{c-function} was introduced in \cite{Chu:2019uoh}, further studied in \cite{Arefeva:2020uec,Hoyos:2020zeg}, and it has already passed various non trivial tests within the holographic scenario. Therefore, we will use it here as a probe for the TQPT.\\
\begin{figure}
    \centering
    \includegraphics[width=0.95\linewidth]{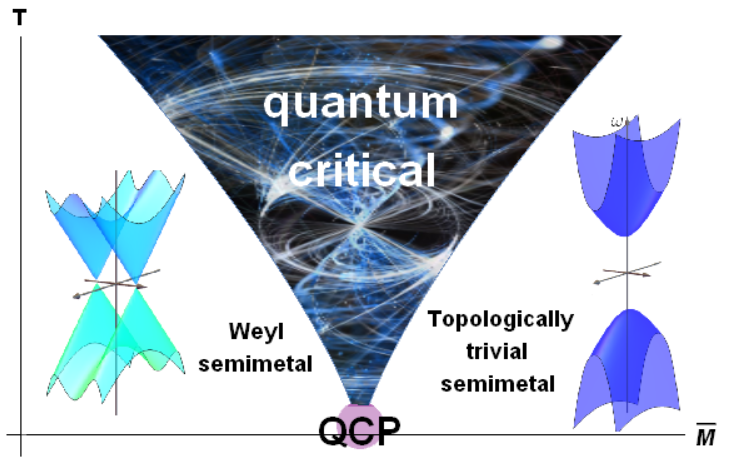}
    \caption{A sketch of the TQPT considered in this work. For our choice of parameters, the transition appears at a critical value $\bar{M}_c \sim 0.744$ between a topologically trivial gapped state and a Weyl semimetal phase.}
    \label{fig:0}
\end{figure}

As a concrete scenario, we consider the holographic Weyl semimetal model introduced in \cite{LANDSTEINER2016453,PhysRevLett.116.081602} (see \cite{Landsteiner:2019kxb} for more details). This setup realizes a quantum phase transition of topological nature between a Weyl semimetal and an insulating phase (see Fig.\ref{fig:0}). The topological distinction between the two phases is described by a topological invariant which has been computed in the context of probe fermions \cite{Liu:2018djq}. Related to this model there are several holographic studies \cite{Ji:2019pxx,Liu:2018spp,Liu:2018bye,Landsteiner:2016stv, Grignani:2016wyz,Copetti:2016ewq,Baggioli:2018afg,Ammon:2018wzb,Ammon:2016mwa}.\\

More broadly, Weyl semimetals (WS) are new 3D materials whose band structure is characterized by singularity points at which the two bands touch, producing linearly dispersing cones \cite{doi:10.1146/annurev-conmatphys-031113-133841}. The low-energy description at those points displays emergent relativistic symmetry and it is described by chiral Weyl spinors always appearing in pairs \cite{NIELSEN1983389}. WS exhibit exotic transport properties which are a direct consequence of quantum field theory anomalies \cite{Landsteiner:2016led}. To comprehend the fundamental dynamics of WS and the TQPT, it is sufficient to consider a simple weakly coupled field theory whose fermionic lagrangian reads \cite{Colladay:1998fq}
\begin{equation}
    \mathcal{L}~=~\bar{\Psi}~\left(i\,\slashed{\partial}\,-\,e\,\slashed{A}\,-\,\gamma_5~\Vec{\gamma}\cdot \Vec{b}~+~M\right)\,\Psi~, \label{fieldtheory}
\end{equation}
where $e$ is the EM coupling, $\gamma_i$ the Dirac matrices,  $A$ is the electromagnetic potential, $M$ a mass parameter and $\Vec{b}$ a vector which describes the separation in momentum space of the two Weyl cones. The system considered exhibits a spectrum which, as expected, depends on the dimensionless ratio $M/|\Vec{b}|$. In the regime $M>|\Vec{b}|$ the system is gapped and the effective fermionic excitations have an effective mass $M_{eff}^2=M^2-|\Vec{b}|^2$, while in the opposite scenario, $M<|\Vec{b}|$, the spectrum is characterized by band inversion and the fermions at the crossing points are massless and separated by the effective parameter $|\Vec{b}_{eff}|^2=b^2-M^2$. Importantly, the axial anomaly implies an anomalous hall conductivity \cite{PhysRevLett.93.206602}:
\begin{equation}
    \sigma_{AHE}\,=\,\frac{1}{(2\pi)^2}\,|\Vec{b}_{eff}|~,
\end{equation}
which is non-zero only in the topological Weyl semimetal phase.\\

In this work, we examine the holographic Weyl semimetal model and we show that the c-function serves as a very efficient tool to diagnostic the location of the TQPT. More generally, it is natural to expect that this concept can apply beyond the realm of holography and could provide a new and fundamental tool for quantum phase transitions evading the standard Landau logic.\\

\section{The topological phase transition}
Our holographic model is defined by the following $5-$dimensional bulk action \cite{PhysRevLett.116.081602}:
\begin{align}
&\mathcal{S}\,=\,\int d^5x~\sqrt{-g}\,\Big[R\,+~12~-\,\frac{1}{4}~F^2\,-\,\frac{1}{4}\,F_5^2 -\,V\left(\Phi\right)\nonumber\\
&-\left(D_\mu \Phi\right)^{*}\left(D^\mu \Phi\right)+\frac{\alpha}{3}\epsilon^{\mu\nu\rho\sigma\tau}A_\nu\Big({F^5}_{\nu\rho}{F^5}_{\sigma\tau}+ 3~F_{\nu\rho}F_{\sigma\tau}\Big)\Big]\label{action}~,
\end{align}
written in terms of a vector $U(1)_v$ field $B_\mu$ with field-strength $F\equiv dB$, an axial vector field $A_\mu$ with field-strength $F_5\equiv dA$ and a complex scalar field $\Phi$ charged under the gauge symmetry $U(1)_v$. Moreover, the covariant derivative is defined as $D_\mu \Phi=\partial_\mu -i q A_\mu \Phi$, and the scalar potential is chosen to be $V(\Phi)=m^2|\Phi|^2+\frac{\lambda}{2}|\Phi|^4$.\\
We use the following anisotropic, in the $x_3$ direction, ansatz for the various bulk fields
\begin{align}
&ds^2 = - f(r) dt^2 + \frac{dr^2}{f(r)} + g(r) \left( dx_1^2 + dx_2^2 \right) + h(r) \,dx_3^2 \, ,\nonumber\\
&A\,=\,A_3(r)\,dx_3\,,\quad \Phi=\phi(r)\,,
\label{ansatz}
\end{align}
where $f(r),~g(r)$ and $h(r)$ depend solely on the radial coordinate $r$. \color{black} We consider asymptotically anti de Sitter configurations for which $f,g,h \sim r^2$ close to the boundary located at $r=\infty$. \color{black} We choose $m^2=-3$ to fix the dimension of the scalar operator dual to the bulk field $\Phi$ to be $\Delta_\mathcal{O}=3$. For this choice, the asymptotics of the gauge field and the bulk scalar are given by:
\begin{equation}
\lim_{r \rightarrow \infty} \, r \,\Phi = M \quad , \quad \lim_{r \rightarrow \infty} \, A_3 = b~ ,
\end{equation}
where $M$ and $b$ are free parameters of the model, which play the same role as those in Eq.\eqref{fieldtheory}. Moreover, without loss of generality, we choose $\lambda=1/10$ and $q=1$.\\
The theory therefore is characterized by two dimensionless parameters taken as $\bar{T}\equiv T/b$ and $\bar{M}\equiv  M/b$ and exhibits a quantum critical transition at $\bar{M}_c\sim 0.74$.\\
At zero temperature, our model admits three types of solutions -- (I) for $\bar{M} > \bar{M}_c$:  an insulating background, (II) for $\bar{M} = \bar{M}_c$: a critical background, and (III) for $\bar{M} < \bar{M}_c$ a semimetal background.

\begin{figure}
    \centering
    \includegraphics[width=0.95\linewidth]{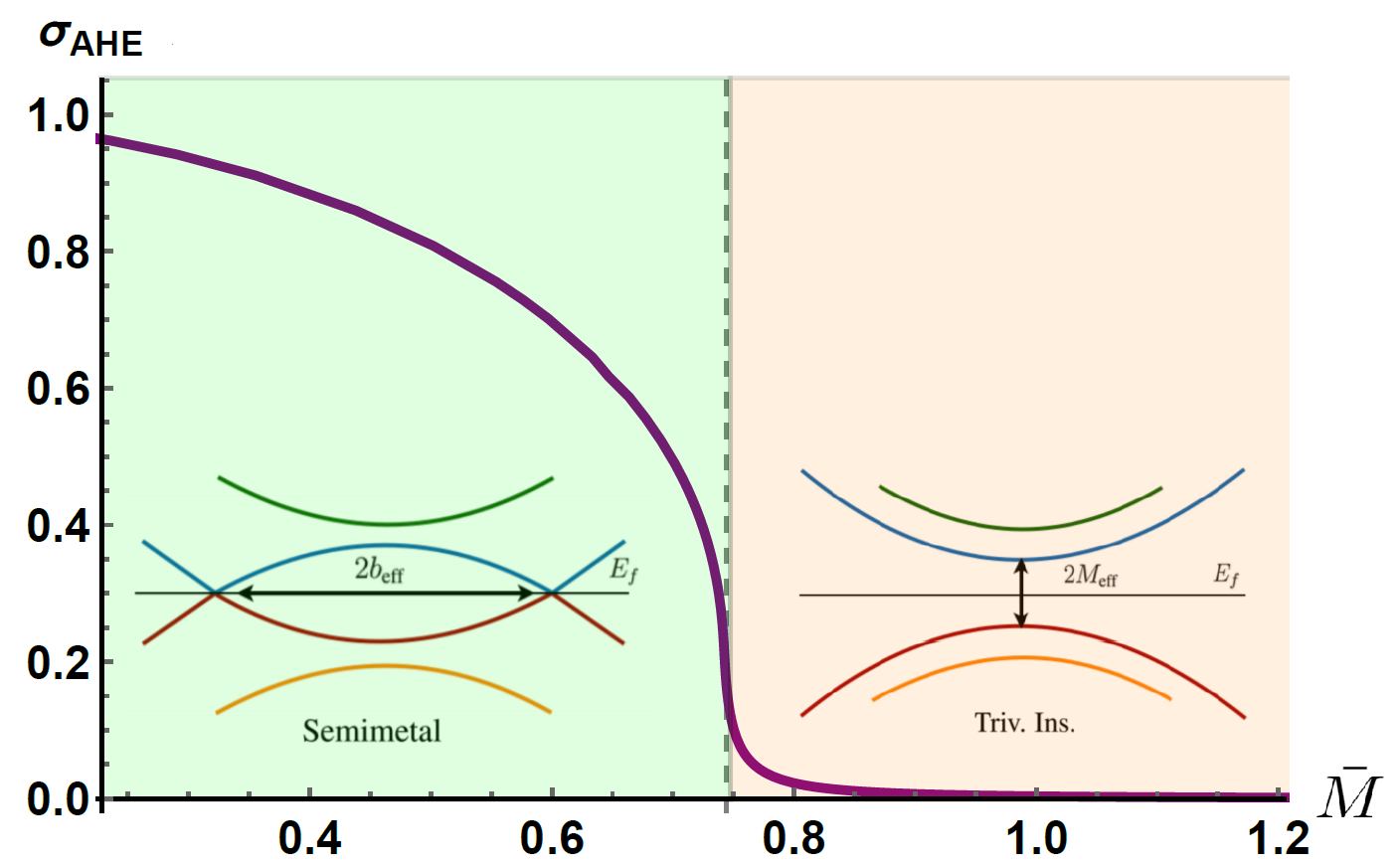}
    \caption{The anomalous conductivity at low temperature $\bar{T}=0.005$ in function of the external parameter $\bar{M}$. The dashed line indicates the position of the quantum critical point $\bar{M}_c \sim 0.744$. The inset displays the topological band crossing which characterizes the semimetal phase. $E_f$ is the Fermi energy of the system.}
    \label{fig:1}
\end{figure}

The full background of the RG flow, can be found only numerically and exhibits different IR fixed points depending on the dimensionless parameter $\bar{M}$. The near-horizon geometry of the topologically trivial gapped solution (I) is an AdS$_5$ domain-wall with $A_3(\rho) = a_1 \,\rho^{\beta_1}, \phi(\rho) = \sqrt{3/\lambda} + \phi_1\, \rho^{\beta_2}$, where the exponents $\beta_{1,2}$ are functions of the parameters $(m,\lambda,q)$ and $\rho$ a new radial coordinate (different from the $r$ used at finite $T$). In this regime (I), the near-horizon value of $A_3$ is always zero, and that of $\phi$ is $\sqrt{3/\lambda}\,$. At the quantum critical point (II), the theory displays an anistropic Lifshitz-like scaling parametrized by $z$, and
induced by the source of the axial gauge field $A_3$. The background can be expressed as
\begin{align}
&f(\rho) = f_0 \,\rho^2 \, , \,\, h(\rho) = h_0 \,\rho^{2/z} \,, \,A_3(\rho) = \rho^{1/z} \, , \,\, \phi(\rho) = \phi_0\,,
\label{bkg:Lifs}
\end{align}
where all the parameters are fixed completely by the choice of $(m,\lambda,q)$. In particular, the anisotropic exponent is given by $z=-(m^2 + \lambda \phi_0^2 - 2q^2)/2q^2$ and takes a value around $z\simeq 2.46$ for our choice of parameters. Null Energy Conditions, the regularity of the solution and the thermodynamic stability generally imply that $z \geq 1$ \cite{Hoyos:2010at,Chu:2019uoh}. 
The near-horizon value of $A_3$ at criticality is always zero, whereas that of $\phi$ is finite equal to $\phi_0$. Finally, by reducing the parameter $\bar{M}$ to values lower than the critical one, we enter in the semimetal phase (III) where the near-horizon geometry is simply AdS$_5$ with  
\be A_3(\rho) = a_1 + \frac{c_1^2 \pi}{4\rho} e^{-\frac{2 a_1}{\rho}},~ \phi(\rho) = \sqrt{\pi} \,\phi_1 \left( \frac{c_1}{\rho} \right)^{3/2}\,e^{- \frac{a_1}{\rho}}\,,
\ee
and the various constants depending on the parameters of the model. In this regime, the near horizon value of $A_3$ is finite, equal to $a_1$; however, $\phi(\rho_0)$ vanishes.\\
To distinguish the two different phases, we consider the anomalous Hall conductivity depicted in Fig.\ref{fig:1}:
\begin{equation}
    \sigma_{AHE}\,\sim\,\,A_3\big|_{\text{horizon}}~.
\end{equation}
The conductivity serves as a non-local order parameter for the TQPT, which vanishes in the topologically trivial insulating phase $(\bar{M}>\bar{M}_c$) and it becomes finite in the Weyl semimetal phase. Interesting, this ''\textit{order parameter}'' does not obey a mean-field theory description but it rather follows a different scaling:
\begin{equation}
    \sigma_{AHE}\,\sim\,\left(\bar{M}_c\,-\,\bar{M}\right)^{0.21}\,,
\end{equation}
which is shown for our lowest temperature in Fig.\ref{fig:1}.
\\

In our scenario, the anisotropy is a characteristic property of the quantum critical point defining the corresponding class of universality, while away of criticality isotropy is always re-emerging. This is a crucial difference with respect to the confinement/deconfinement phase transitions in Einstein-Dilaton-Axion theories \cite{Giataganas:2017koz} which contain a finite degree of anisotropy that remains invariant along the different phases.\\

\section{The c-function for Lifshitz-Like Systems}
For the sake of introducing the notion of the anisotropic c-function, let us temporarily consider an arbitrary dimensional spacetime in which the $d-$dimensional spatial subspace can be decomposed in a transverse and parallel sets with respective dimensions $d_1,d_2$, ($d_1+d_2=d$), enjoying different scalings:
\begin{equation}
  [\parallel]=L^{n_1}~,~[\perp]=L^{n_2} \,,
\end{equation}
and therefore breaking the rotational $SO(d)$ invariance to SO($d_1$) $\times$ SO($d_2$).
The natural proposal for the c-function of these theories is given by \cite{Chu:2019uoh}
\begin{equation}
    \label{cfunction}
c_\parallel := \beta_\parallel\, \frac{l_\parallel^{d_\parallel-1}} {H_\parallel^{d_1-1} H_\perp^{d_2}}\,
\frac{\partial S_\parallel}{\partial \ln l_\parallel}~,\,\,\,c_\perp := \beta_\perp \,\frac{l_\perp^{d_\perp-1}} {H_\parallel^{d_1} H_\perp^{d_2 -1}}\,
\frac{\partial S_\perp}{\partial \ln l_\perp},
\end{equation}
where $S_{\parallel} (S_\perp)$ is the entanglement of the slab geometry with length $l_\parallel (l_\perp)$ along one of the spatial $\parallel, (\perp)$ dimensions and $H$ corresponds to the UV cut-offs. The $d_\parallel$ and $d_\perp$ are the effective dimensions:
\begin{equation}\la{effdim}
    d_\parallel := d_1 + d_2 \frac{n_2}{n_1}\,,\,\,\,d_\perp:= d_1 \frac{n_1}{n_2} +d_2~,
\end{equation}
of the two rotational invariant spatial planes of dimensions $d_1$ and $d_2$, where the initial isotropy has been broken to. The parameters $\beta_{\parallel,\perp}$ are just dimensionless normalization constants. The entangling surface in $(\parallel,\perp)$ directions is computed holographically with the usual strategy on anisotropic probes introduced in \cite{Giataganas:2012zy}. When the c-function is computed at a certain fixed point, the effective dimensions are identified with the corresponding scaling exponents. Importantly, the above definition Eq.\eqref{cfunction} reduces to the conformal and isotropic c-function  \cite{Casini:2004bw,Casini:2006es,Ryu:2006ef} when the symmetries (in this case isotropy) are restored.

\section{Uncovering the quantum critical point}
To uncover the criticality of the theory, we firstly obtain the numerical background of Eq.\eq{ansatz} for $\bar{M} \in [0,4]$, while keeping fixed the dimensionless temperature $\bar{T}$. We are primarily interested in the zero temperature and extremely low temperatures ($\bar{T}\simeq 0.005$). Nevertheless, we have directly checked that similar results are obtained for slightly larger values ($\bar{T}=0.05,0.1$).\\

In order to locate the quantum critical point, we compute the c-function corresponding to an entangling surface with fixed large enough boundary length to extend away from the boundary into the bulk. This type of entangling region probes the deep IR and provides good accuracy on locating the phase transitions. The c-function defined in \eqref{cfunction} has the advantage of being valid and well-defined even for anisotropic IR phases. Exploiting this feature, we are able to compute it  across the full  phase diagram $(\bar{M},\bar{T})$. Notice that this would have been impossible by using of the isotropic c-function, because the quantum critical point exhibits a strong anisotropic character.\\

Practically, the external parameter $\bar{M}$ is dialed in a range able to cover the three different phases of the theory: trivial insulator ($\bar{M} > 0.744$), critical point ($\bar{M} \sim 0.744$) and Weyl semimetal ($\bar{M} < 0.744$) (see Figures \ref{fig:0} and \ref{fig:1}). The c-function develops a clear pattern. As we approach the quantum critical point, it increases and it reaches a maximum exactly at the quantum critical point with the Lifshitz-like symmetry. In this sense, the c-function acts as a very accurate probe to locate the topological quantum critical point.
\\
\begin{figure}[t]
    \centering
    \includegraphics[width=0.85\linewidth]{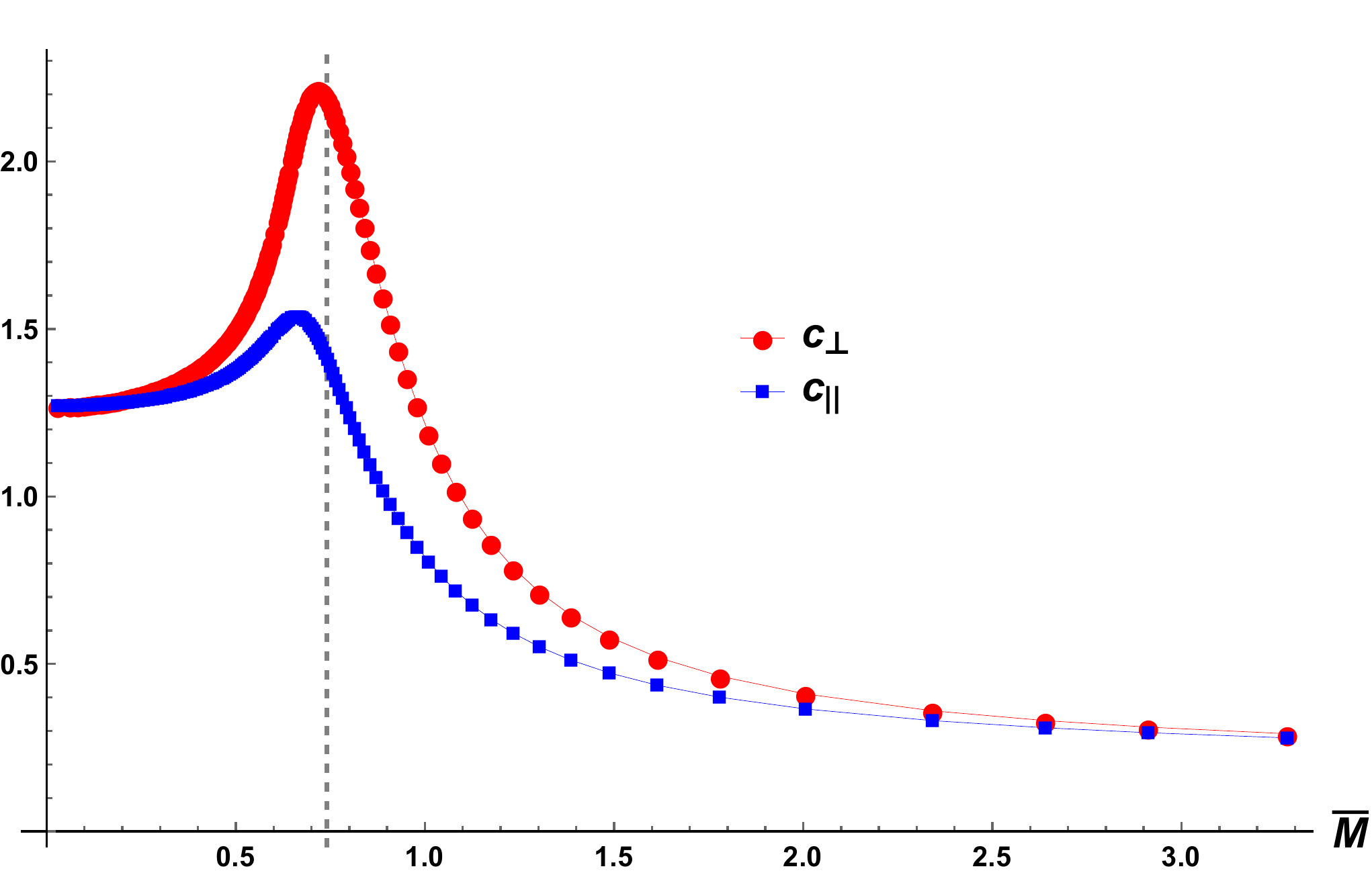}
    \caption{  The parallel and transverse c-functions at $\bar{T}=0.005$ in function of the external parameter $\bar{M}$. The dashed line indicates the position of the quantum critical point, $\bar{M}_c \sim 0.744$.
    Both functions have normalized magnitudes for presentation reasons.
    }
    \label{fig:c_function}
\end{figure}

\begin{figure}
    \centering
    \includegraphics[width=0.9 \linewidth]{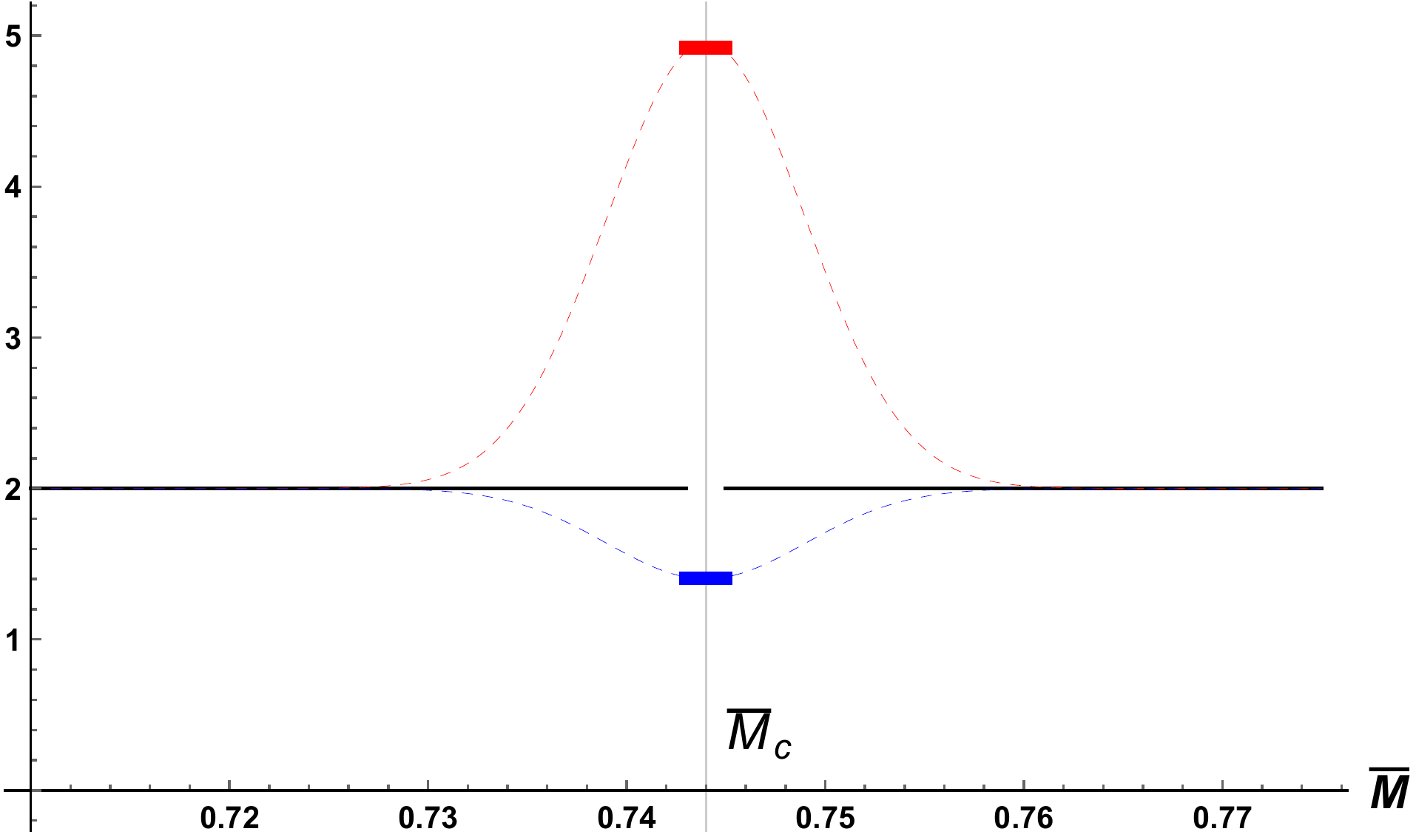}
    \caption{\color{black} The normalized values of the c-functions at $T=0$ in function of $\bar{M}$. The black lines indicate the AdS value $c/\beta(d)=2$, normalized with $\beta$. The red and blue line displays the value of $c_\perp/\beta_\perp(d_\perp),c_\parallel/\beta_\parallel(d_\parallel)$ at the quantum critical point where the c-function is clearly discontinuous.  The normalizations are for presentation reasons, since our main focus is the position and the existence of a saddle point of the c-function  at the phase transition. \color{black}}
    \label{fig:zero}
\end{figure}

\color{black}
\color{black} The $T=0$ near-horizon geometry of our background is an anisotropic Lifshitz-like space-time. The c-function is computable analytically in this type exact geometries that maintain the anisotropy along the RG flow. \color{black}By a straightforward application of the formulas of the Appendix B and eq. \eq{ent_all}, we obtain for the cut-off independent terms:\color{black}
\bea\la{eeen11}
&&S_\perp\simeq -N^2 H_\perp H_\parallel  \frac{\beta_\perp{}^{-1}}{l_x^{1+\frac{1}{z}}}\,,\\
&&S_\parallel\simeq -N^2 H_\perp^2 \frac{\beta_\parallel{}^{-1}}{l_y^{2 z}}\,, \la{eeen22}
\eea\color{black}
where the $\beta_\perp$ and $\beta_\parallel$ 
\color{black}are constants that depend on the effective dimensions $d_\perp$ and $d_\parallel$ respectively through Gamma functions. 
\color{black} Since we focus on surfaces corresponding to large entangling distances $l$, we may use  for the purpose of obtaining the analytical results approximately the equations \eq{eeen11} and \eq{eeen22}, and  the effective dimensions are specified to the ones of the IR regime as we do later in our numerical methods.  \color{black} By extracting the c-function from the entanglement entropy, we obtain $c_\perp=\beta_\perp{}^{-1}\prt{1+1/z}$ and $c_\parallel=\beta_\parallel{}^{-1}2 z$ with an appropriate normalization. The two functions $c_\parallel$ and $c_\perp$ converge for $z=1$ to a single value since also $\beta_\parallel=\beta_\perp$. Away of the quantum critical point, the constants reflect the AdS nature of the fixed points for $\bar{M}\gtrless \bar{M}_c$. At the quantum critical point  $\bar{M}_c$, the symmetries of the system change, resulting into a separation of the c-functions $c_\perp\neq c_\parallel$  and a discontinuous jump in their values 
to new constants, each one depending on the corresponding effective dimensions  $(d_\perp,d_\parallel)=(2+1/z,~1+2 z)$ introduced in \eq{effdim} (see Fig.\ref{fig:zero}). This discontinuity is the characteristic signal of the quantum phase transition at zero temperature via the c-function.\color{black}

\color{black}
The separation of the two c-functions at the critical point, is the signal of the a fast spatial re-organization of the degrees of freedom due to the emergent anisotropy of the system at criticality and it is confirmed by our zero temperature results where thermal effects are negligible.
\color{black}
\\

At finite and low temperatures, within the so-called quantum critical region depicted in Fig.\ref{fig:0}, this discontinuous feature is smoothed out by thermal effects and the task of identifying the crossover point with the c-function is more demanding. We show our main results in Fig.\ref{fig:c_function}. \color{black} For large entangling surfaces both c-functions detect very accurately the position of the anisotropic quantum critical point. Despite the finite temperature effects, the finite $T$ c-function retains a memory of the $T=0$ critical point. \color{black} \color{black}This is a common feature reminiscent of the so-called quantum critical region which might make the experimental measurements (which are definitely impossible at zero temperature) more feasible. \color{black}
\color{black} A technical question that one may ask is how do we identify effective dimensions from the numerical data.  We isolate the critical exponent by computing the derivative of the logarithmic ratio $g_{11}/g_{33}$ at fixed radial distance and taking into account that we may always re-scale the transverse metric element to have a known isotropic scaling.

Since our comparison involves shifting the parameter $\bar{M}$, there is a question on the way we normalize the rest of the scales. This is relevant to our computation since in principle we like to probe the theory at a certain energy  as $\bar{M}$ changes. We notice that most qualitative details of the c-functions are weakly dependent on the comparison scheme while the derivatives of the entanglement are strongly dependent.  A choice of scheme is to keep fixed  the length of the entangling region to a constant value, where the holographic entangling surface turning point varies with $\bar{M}$. Alternatively, we may normalize the length of the entangling surface with respect to the energy scale of the gravity theory and maintain this dimensionless quantity constant  while  the surface's turning point and length varies with respect to $\bar{M}$. Another option is keep constant the turning point of the holographic surface versus the proximity of the black hole horizon as $\bar{M}$ varies. Irrespective of the comparison scheme, we obtain a clear signal of the phase transition for the large entangling surfaces we study in this paper since they  probe the IR deeply. The transverse c-function, nicely develops always a maximum located at the critical point. The parallel one is more sensitive to the scheme especially for small entangling surfaces. It still indicates the location of the critical point but it does not always develop a maximum around it, while in that case still one of its derivatives develops a zero value.
\color{black}
\\
\begin{figure}[t]
    \centering
    \includegraphics[width=0.95\linewidth]{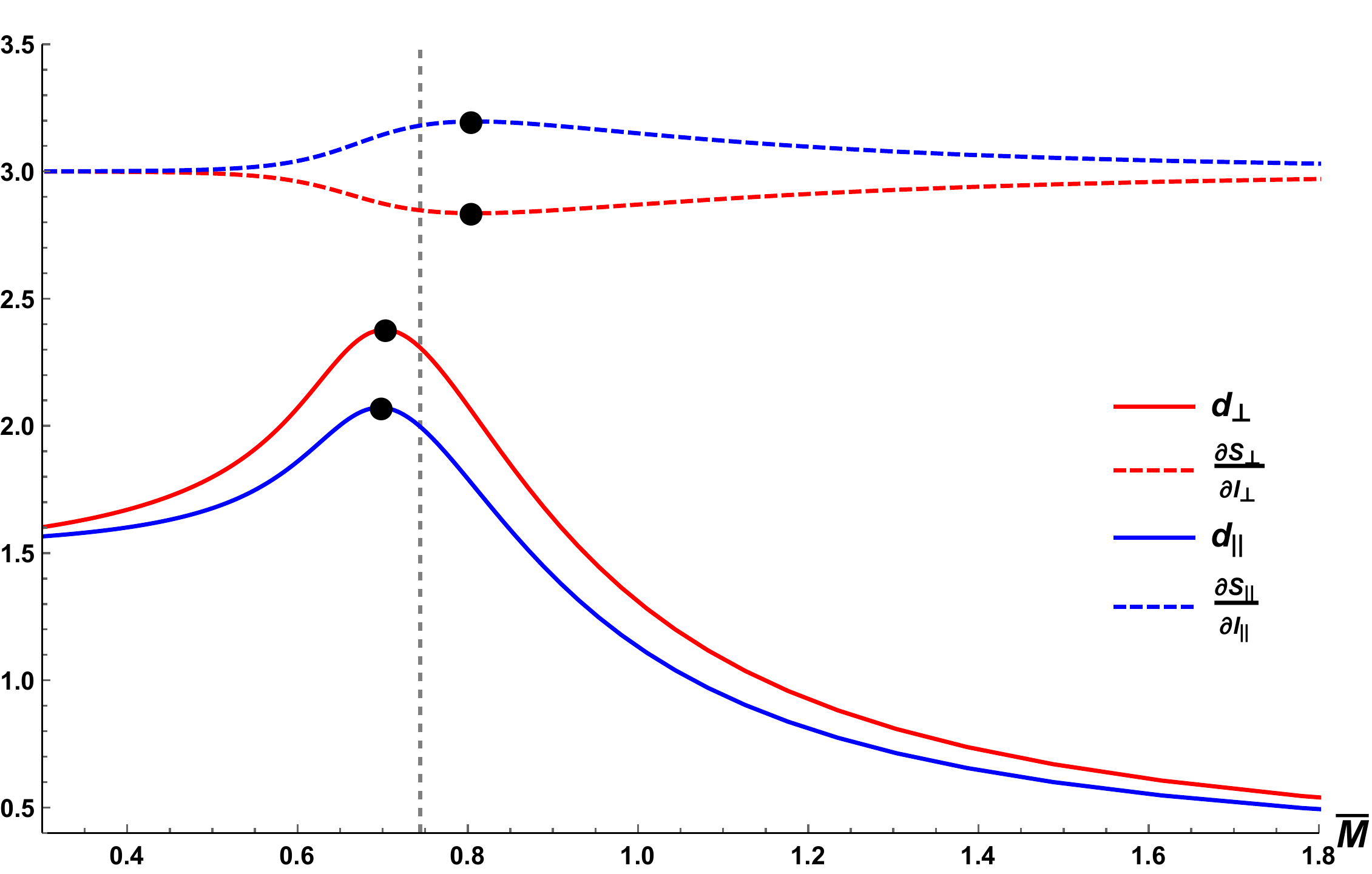}

    \caption{The effective dimensions $d_{\parallel,\perp}$ and the EE derivatives $\partial S_{\parallel,\perp}/\partial l_{\parallel,\perp}$ in function of the external parameter $\bar{M}$, around the QCP. The black dots indicate the position of the saddle points while the vertical line the location of the QCP. Notice that the effective dimensions overestimate the fixed point while the derivatives underestimate it. The c-function saddle point takes into account both. Note that for low and large $\bar{M}$ the quantities converge. The comparison scheme used here is for constant entangling length as $\bar{M}$ varies and the magnitude normalizations are fixed for presentation reasons.}
    \label{fig:dim}
\end{figure}

At this point, few comments are in order. The c-function defined in \eqref{cfunction} contains accurate information to signal the phase transition in its terms. In fact, the information of the phase transition is included in the effective dimensions $d_\parallel$ and $d_\perp$ since the critical point has Lifshitz-scaling anisotropy, in contrast to the AdS phases. Information of the phase transition is also included in the entangling surface itself. We show the behaviour of the effective dimensions and the EE derivatives\footnote{Different EE derivatives have been already considered in \cite{Ling:2016wyr,Ling:2015dma} for certain holographic Q-lattices model.} in function of $\bar{M}$ in Fig.\ref{fig:dim}. The first observable over-estimates the location of the QCP, at least for our chosen values\footnote{The parallel effective dimension displays a minimum at the QCP instead of a maximum. This is similar to the behaviours of the conductivities, viscosities and butterfly velocities \cite{Baggioli:2018afg}.}, while the second under-estimates it. Interestingly, the exact combination of the two, which appears in \eqref{cfunction}, is the one that pinpoints the precise position of the quantum critical point with the greatest accuracy. 

\color{black}
Moreover, a larger entangling surface implies that the corresponding surface probes deeper the IR structure of the theory. In this regime the saddle points of the derivatives of the entanglement entropy are approaching closer to the points of the phase transition.  For larger entangling regions at the boundary, the entanglement entropy received contributions from the thermal entropy. This is a known statement in thermal theories e.g. \cite{Giataganas:2019wkd}. We point out that, the thermal contributions on the computed entanglement entropy, do not affect the ability of the c-function to locate the probe. In fact the thermal entropy or any non-local observable would give a hint of the phase transition in systems of finite temperature. Notice also that our approach is still valid for smaller entangling surfaces where the $c_\perp$ always develops a saddle point at the critical regime, while the $c_\parallel$ may develop an anomalous conductivity  like behavior, similar to the one depicted in Fig. \ref{fig:1} and the phase transition signal then comes by the peak that its derivative develops. As we have stated already in this paper we work with large entangling surfaces, where both the $c$-functions signaling accurately the QCP, i.e. as in Fig. \ref{fig:c_function}.\\ 
\color{black}



\begin{figure}
    \centering
    \includegraphics[width=0.9\linewidth]{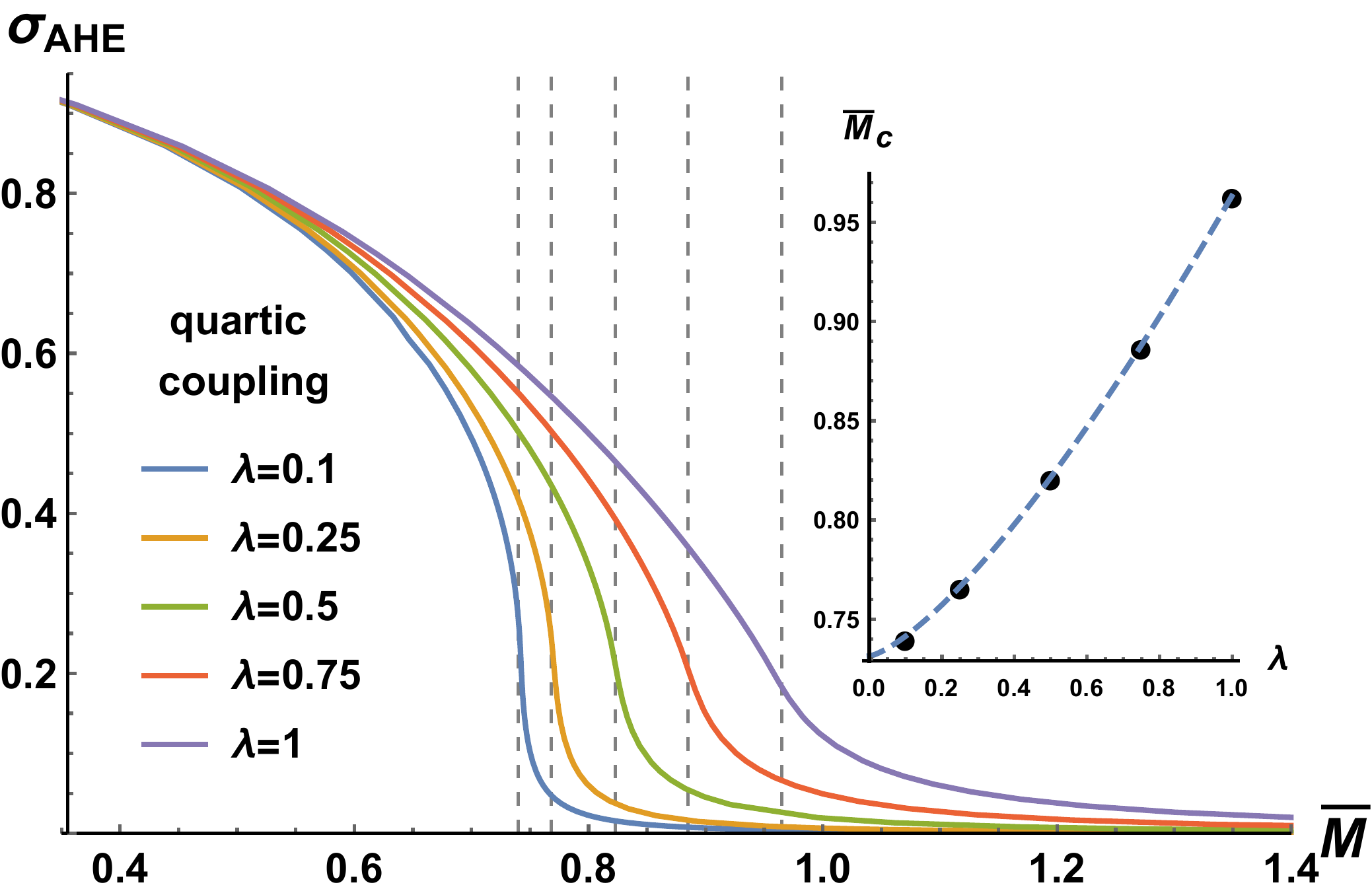}
    \caption{\color{black}The anomalous conductivity at low temperature with respect to quartic coupling, the dashed lines denote the positions of the QCP. In the small plot the dots correspond the QCP with respect to the quartic couplings.\color{black} }
    \label{fig:new1}
\end{figure}

\begin{figure}
    \centering
    \includegraphics[width=0.9\linewidth]{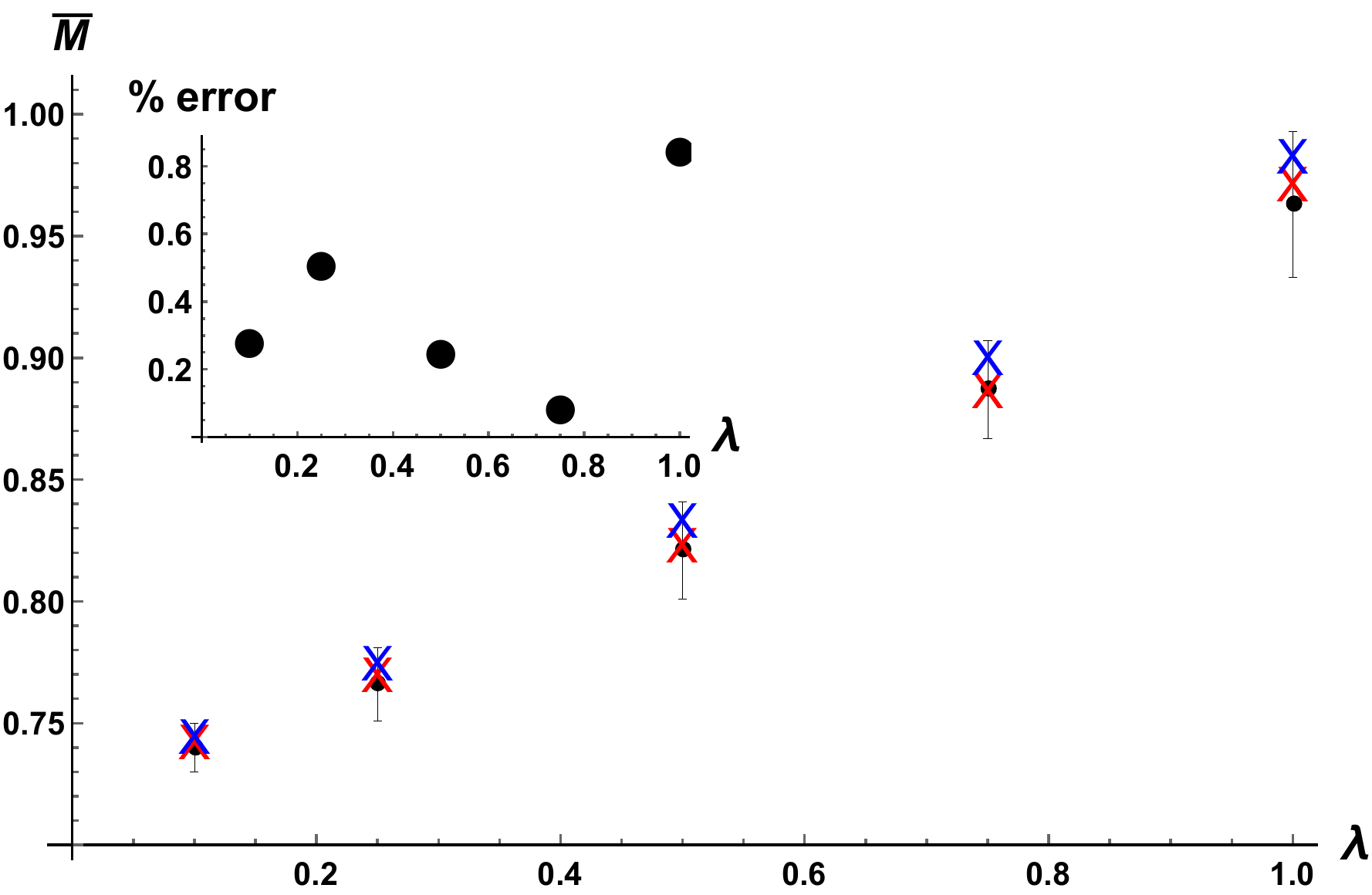}
    \caption{\color{black}The $X$ marks denote QCP computed by the $(c_\perp,c_\parallel)$-functions for a varying quartic coupling. The  dots are the QCP of the system and the error bars highlight the \color{black}uncertainty  of their identification due to thermal effects. The error bars are estimated using the value at which the anomalous hall conductivity approaches the benchmark value $\sigma_{AHE}=0.1$. The inset highlights the difference between the actual location of the QCP and the estimate of our probe.\color{black} The c-functions pinpoint the QCP with great accuracy.\color{black} }
    \label{fig:new2}
\end{figure}

\color{black}In order to confirm more robustly the ability of the c-function to detect the quantum critical point, we are performing a more detailed analysis on the theory parameters by changing the quartic coupling of the model $\lambda.$ By increasing it, the critical point $\bar{M}_c$ gets larger as shown in Fig. \ref{fig:new1}. Then we compute the saddle points of the c-functions and in in Fig.\ref{fig:new2} we show the comparison between the location of the critical points for various $\lambda$ and the positions of the saddle points of the c-functions . The c-functions are able to locate the quantum critical point with an error $< 1 \%$. This is a further evidence the universality of our findings. \color{black} 

\section{Discussion}
In this work we have considered a holographic model exhibiting a topological quantum phase transition (TQPT). The quantum critical point, which is related to the transition between a topologically trivial insulator and a gapless Weyl semimetal, displays a Lifshitz-like anisotropic critical point and critical scalings not compatible with the standard Landau paradigm. We have shown that the anisotropic c-function attains a universal maximum at the location of the quantum critical point and as such it serves as a very accurate and efficient probe to detect the TQPT.\\

\color{black} Our probe is very successful in locating the critical point despite its topological nature, nevertheless, given an unknown phase transition it is definitely not able to capture or detect whether it is topological or not. It would be interesting to investigate more refined probes able to discern the topological information of the critical point. A natural possibility would be to consider the so-called \textit{topological entanglement entropy}, which has been already discussed in holography in \cite{Pakman:2008ui}.\\

Moreover, it is imperative to understand better, beyond the qualitative argument of the dofs counting, what is the fundamental origin of the c-function peaks and if it is connected or not with the features of the transport coefficients discussed in \cite{PhysRevLett.116.081602,Landsteiner:2016stv}. We leave these questions for the near future.\color{black}

\subsection{Acknowledgments}
We would like to thank C-S Chu, S.Cremonini, J.P. Derendinger, M.Flory, S.Grieninger, L.Li, K.Landsteiner, B.Padhi, Y.Liu, and Y.W.Sun for useful discussions and suggestions. M.B. acknowledges the support of the Spanish MINECO’s ``Centro de Excelencia Severo Ochoa'' Programme under grant SEV-2012-0249.  D.G. research has been funded by the Hellenic Foundation for Research and Innovation (HFRI)
and the General Secretariat for Research and Technology (GSRT), under grant agreement No
2344.

\bibliographystyle{apsrev4-1}
\bibliography{cfunction}
\appendix
\onecolumngrid
\section{Appendix A: Background} \label{app1}
Plugging our background ansatz into the bulk action, we obtain the following equations of motion:
\bea\label{backEOMs1}
&&    f'' +\frac{h'}{2 \,h}\, f'\, - f\left(\frac{g''}{g}+\frac{g'\, h'}{2 \,g\,
   h}\right)=\,0\,,\\
 &&   \frac{f''}{2
   f} +\frac{g''}{g}+\frac{g'\, f'}{g\,f}-\frac{{g'}^2}{4
   g^2}-\frac{{A_3'}^2}{4 \,h}+\frac{1}{2} {\phi
   '}^2 + \frac{m^2 \phi^2}{2 f}-\frac{q^2 A_3^2 \phi^2}{2 h
   f}+\frac{\lambda \phi^4}{4 f}-\frac{6}{f}\,=\,0\,, \\
&&  \frac{1}{2} {\phi '}^2+   \frac{{A_3'}^2}{4 h} - \left(\frac{g'}{2 g\,f}+\frac{h'}{4 \,h\,
   f}\right)f'-\frac{g'\, h'}{2 g\, h}-\frac{{g'}^2}{4
   g^2}- \left({m^2}+\frac{q^2 A_3^2 }{h}+\frac{\lambda  \phi^2}{2} \right) \frac{\phi^2}{2f}+\frac{6}{f}=0 \, ,\\
&&    \phi ''+\phi '
   \left(\frac{g'}{g}+\frac{h'}{2
   h}+\frac{f'}{f}\right)-\frac{\lambda  \phi^3}{f} - \left(\frac{q^2\,A_3^2}{h\,f}+\frac{m^2}{f}\right) \phi =0~,\\
&& A_3''+A_3'\prt{\frac{g'}{g}-\frac{h'}{2 h}+\frac{f'}{f}}-\frac{2 q^2 A_3 \phi^2}{f}=0~.
\eea
At the UV boundary ($r=\infty$) the asymptotic expansion of the bulk fields is given by:
\begin{align}
&f\,=\,r^2\,+\,\dots\,,\quad g\,=\,r^2\,+\,\dots\,,\quad h\,=\,r^2\,+\,\dots\,,\quad A_3\,=\,b\,+\,\dots\,,\quad \phi\,=\,\frac{M}{r}\,+\,\dots\,.
\end{align}
Our theory has the following three scaling symmetries
\bea
&&  (x_1,x_2)\rightarrow a(x_1, x_2),\,\,~~g\rightarrow g a^{-2}~  ;\qquad    x_3\rightarrow a x_3,\,\,~~h \rightarrow a^{-2}h,\,\,A_3\rightarrow A_3 a^{-1} \\
&& r\rightarrow a r,\,\,(t,x_1,x_2,x_3)\rightarrow (t,x_1,x_2,x_3)/a,\,\,~~(f,g,h)\rightarrow a^2(f,g,h),\,\,A_3 \rightarrow a A_3~,
\eea
which are used to rescale the coefficients of the three different metric functions $(f,g,h)$ at the boundary to unity. This is why the boundary field theory depends only on the parameters, $T, b, M$ which can be reorganized into two dimensionless quantities  $\bar{T}\equiv T/b$ and $\bar{M}\equiv M/b$.\\
Approaching the black-brane horizon ($r_h$), the expansion for the bulk fields can be written as
\begin{align}
\begin{split}
&f \simeq 4\,\pi\, T\,(r-r_h)+f_2\,(r-r_h) \,\,,\,\,g\simeq g_1+g_2\,(r-r_h)\,\,,\,\,\, h\simeq h_1+h_2\,(r-r_h)\,\,, \\
&A_3\simeq {A_3}^{(1)}+{A_3}^{(2)}\,(r-r_h) \,\,,\,\, r \phi \simeq \phi_1\,+\,\phi_2\,(r-r_h) \, .
\end{split}
\end{align}
$A_{3}^{(1)}$ and $\phi_1$ are the only free parameters, controlled by the boundary data $\bar{T}$ and $\bar{M}$. 
In summary, we can reduce the number of the independent horizon parameters $(T,r_h,g_1,h_1,{A_3}^{(1)},\phi_1)$ to $(T,{A_3}^{(1)},\phi_1)$ using the above scaling symmetries. At the conformal boundary, these parameters can be mapped into the triplet $(T, M, b)$. Following this procedure, we obtain numerically our background using the shooting method  with respects to parameters $\bar{T}$ and $\bar{M}$.

\color{black}
\section{Appendix B: A short derivation of the Entanglement Entropy}\label{app2}

Let us provide a walkthrough for the entangelement formulas in generality for the anisotropic theories.  We consider a 5-dim holographic spacetime 
\be 
ds^2=-g_{00}(u) dt^2+g_{11}(u) dx_1^2+ g_{22}(u) dx_2^2+g_{33}(u) dx_3^2+ g_{rr}(u) du^2~, 
\ee
with boundary, lets say, at $u=0$. We consider the subsystem cut out along the $x_1$-direction and of length $l_1$. We follow the notation of \cite{Chu:2016pea}, while further the details the effective dimensions of the entanglement entropy in anisotropic theories are given in \cite{Chu:2019uoh}. The minimal surface ending on the boundary of the subsystem is given by
\be 
S= H_2 H_3 \int_0^{l_1} d\s A \sqrt{g_{11}+g_{rr} u(\s)'{}^2} ~,
\ee
where $A:= \sqrt{g_{22} g_{33}}$ and $H$ is the infrared regulator. The equations of motion read
\be 
u'{}^2=\frac{g_{11}\prt{A^2 g_{11}-c^2}}{c^2 g_{rr}}~,
\ee
where $c=A(u_*) \sqrt{g_{11}(u_*)}$ with $u_*$ the turning point. The length of the entangling region is related to the turning point of the surface by
\be 
l_1=2 \int_0^{u_*} du \frac{c^2 g_{rr}}{g_{11}\prt{A^2 g_{11}-c^2}}~
\ee
and the entanglement is given by
\be \la{ent_all}
S= 2 H_2 H_3 \int_0^{u_*} dr A \sqrt{\frac{g_{rr}}{1-\frac{c^2}{A^2 g_{11}}}}~.
\ee
\color{black}

\end{document}